\documentclass{amsart}
\usepackage{amssymb,amsmath,amscd, amsthm,stmaryrd,verbatim, mathrsfs}

\newtheorem{theorem}{Theorem}[section]

\theoremstyle{definition}

\theoremstyle{remark}

\numberwithin{equation}{section}

\newcommand{\bb}{\mathbb}

\newcommand{\mr}{\mathrm}
\newcommand{\frk}{\mathfrak}

\usepackage{hyperref}
\begin{document}

\title{The Universal Kepler Problem}
\author{Guowu Meng}
\address{Department of Mathematics, Hong Kong Univ. of Sci. and
Tech., Clear Water Bay, Kowloon, Hong Kong}


\email{mameng@ust.hk}
\thanks{The author was supported by Qiu Shi Science and Technologies Foundation while he was a member at the Institute for Advanced Studies in the cadmic year 2010-2011. He was also supported by the Hong Hong Research Grants Council under RGC Project No. 16304014.}

\keywords{Kepler Problem, Laplace-Runge-Lenz Vector, Euclidean Jordan Algebra, TKK Algebra}

\maketitle
\begin{abstract}
For each simple euclidean Jordan algebra $V$, we introduce the analogue of hamiltonian, angular momentum and Laplace-Runge-Lenz vector in the Kepler problem. Being referred to as the universal hamiltonian, universal angular momentum and universal Laplace-Runge-Lenz vector respectively, they are elements in (essentially) the TKK (Tits-Kantor-Koecher) algebra of $V$ and satisfy commutation relations similar to the ones for the hamiltonian, angular momentum and Laplace-Runge-Lenz vector in the Kepler problem. We also give some examples of Poisson realization of the TKK algebra, along with the resulting classical generalized Kepler problems. For the simplest simple euclidean Jordan algebra (i.e., $\mathbb R$), we give examples of operator realization for the TKK algebra, along with the resulting quantum generalized Kepler problems.

\end{abstract}


\section {Introduction}
Recall that, in the Kepler problem,  the hamiltonian is
\begin{eqnarray}
{\mathrm H}={1\over 2}{\mathbf p}^2-{1\over r}.
\end{eqnarray}
Here, $r$ is length of $\mathbf r \in \bb R^3_*:=\bb R^3\setminus\{\mathbf 0\}$ and $\mathbf p$ is the (linear) momentum.

The hamiltonian ${\mathrm H}$ is clearly invariant under rotations of $\bb R^3$, thanks to Noether's theorem, the angular momentum
\begin{eqnarray}\mathbf L=\mathbf r \times \mathbf p
\end{eqnarray}
is conserved. 

What is special about the Kepler problem is the existence of an additional conserved quantity, i.e., the Laplace-Runge-Lenz vector

\begin{eqnarray}
\mathbf A=\mathbf L \times \mathbf p+{\mathbf r\over r}.
\end{eqnarray}

Not everyone agrees that this is a proper name for this vector because of the long history of its rediscovery: Jakob Hermann initially discovered it for a special case of the inverse-square central force \cite{Herman}, Johann Bernoulli generalized it to its modern form \cite{Bernoulli} in 1710, and at the end of the 18th century, Pierre-Simon de Laplace rediscovered it analytically \cite{Laplace}. In the literature, this vector is sometimes called the Runge-Lenz-Laplace vector \cite{Yanovski}.

It is well-known that ${\mathrm H}$, $\mathbf L$ and $\mathbf A$ satisfy the following Poisson bracket relations (For the definition of Poisson brackets, please consult Section 5 of Chapter III in Ref. \cite{LM}):

\begin{eqnarray}\label{Lenz}
\begin{array}{ll}
\{{\mathrm H}, L_i\} = 0, & \quad \{{\mathrm H}, A_i\} = 0,\\

\{L_i, L_j\}= \epsilon_{ijk}L_k, & \quad \{L_i, A_j\} =\epsilon_{ijk}A_k,\quad \{A_i, A_j\} = -2{\mathrm H}\epsilon_{ijk}L_k.
\end{array}
\end{eqnarray}
Here, $\epsilon_{ijk}$ is the antisymmetric tensor such that $\epsilon_{123}=1$, and a summation over the repeated index $k$ is assumed in the above.

\vskip 5pt
When passing to the quantum case, nothing is lost. First of all, we have
\begin{eqnarray}
\begin{array}{rcl}
{\mathrm H} &= &-{1\over 2}\Delta-{1\over r},\\
\mathbf L &= & -\mathrm{i}\mathbf r \times \nabla,\quad
\mathbf A =-{\mathrm{i}\over 2}(\mathbf L \times \nabla- \nabla\times \mathbf L)  +{\mathbf r\over r};
\end{array}
\end{eqnarray}
secondly, relation (\ref{Lenz}) still holds when Poisson brackets are replaced by commutators.

\vskip 10pt
The goals of this article are to introduce the analogues of ${\mathrm H}$, $\mathbf L$ and $\mathbf A$ for each simple euclidean Jordan algebra \cite{PJordan33}, derive the analogue of relation (\ref{Lenz}), and demonstrate via examples their relevance to generalized Kepler problems.
To do that, we need to  digress into simple euclidean Jordan algebra and the associated TKK (Tits-Kantor-Koecher) algebra \cite{TKK1}, \cite{TKK2, TKK3}.

\vskip 10pt
\underline{Added on Dec. 1, 2014}. This paper was initially written in the winter of 2010. Since then, Refs. \cite{meng11a, meng11b, meng12} appeared which in one way or another all demonstrate that the Jordan algebra approach to the Kepler problem in relationship to the universal Kepler problem advocated here is quite natural and fruitful. We would like to remark that the universal nature discovered here for the Kepler problem is quite a common trait shared by many beautiful mathematical objects: cohomology groups, vector bundles, characteristic classes, R-matrices, knot invariants of finite type, etc.

\section{TKK Algebra}
Let $V$ be a finite dimensional simple euclidean Jordan algebra. This means that, $V$ is a real vector space of positive dimension; and there is a bilinear map $V\times V\to V$ which maps $(a, b)$ into $ab$ such that, for any $a, b\in V$,  1) $ab=ba$, 2) $(ab)a^2=a(ba^2)$, 3) $a^2+b^2=0\implies a=b=0$; moreover, $V$ has no nontrivial ideal. It turns out that there is a multiplicative unit element $e$ in $V$.

For $a\in V$, we use $L_a$ to denote the multiplication by $a$, so $L_a (b)= ab$. Clearly $L_a$ is an endomorphism on $V$ and is linearly dependent on $a$. We assume the invariant inner product $\langle\mid\rangle$ on $V$ is the inner product such that the unit element $e$ has unit length, i.e., 
$$ \langle a\mid b\rangle:={1\over \dim V}\mbox{Tr }L_{ab}$$
for any $a, b\in V$. Here, the inner product $\langle\mid \rangle$ is invariant means that $L_a$ is self-adjoint with respect to it, i.e., $\langle a b\mid c\rangle = \langle b\mid ac\rangle$ for any $a, b, c\in V$.

We denote the Jordan triple product of $a$, $b$, $c$ by $\{abc\}$. Recall that
$$
\{abc\}:=a(bc)-b(ca)+c(ab).
$$
We denote the endomorphism $c\mapsto \{abc\}$ by $S_{ab}$. It is clear that
$$S_{ab}=[L_a,L_b]+L_{ab}$$ 
and is bilinear in $(a, b)$. It is clear that $L_a=S_{ae}=S_{ea}$. We shall use $S'_{ab}$ to denote the adjoint of $S_{ab}$ with respect to the inner product on $V$. Note that $S_{ab}'=S_{ba}$, and if we identify $V^*$ with $V$ via the invariant inner product, $S_{ab}^*$: $V^*\to V^*$ can be identified with $-S_{ba}'$. 

One can check that
\begin{eqnarray}\label{stralg}
[S_{ab}, S_{cd}]=S_{\{abc\}d}-S_{c\{bad\}},
\end{eqnarray}
so $S_{ab}$'s form a real Lie algebra --- the {\bf structure algebra} $\frk{str}$ of $V$. The commutation relation in Eq. (\ref{stralg}) says that, in $S_{cd}$, $c$ and $d$ behave as a $\frk{str}$-vector and $\frk{str}$-covector respectively.  In general,
$\frk{str}=\frk{str}'\oplus \bb R$, where $\frk{str}'$, a semi-simple real Lie algebra, is called the {\bf reduced structure algebra}.

It is an independent discovery of Tits, Kantor, and Kroecher \cite{TKK1}, \cite{TKK2, TKK3} that the real reductive Lie algebra $\frk{str}$ can be naturally extended to a real simple Lie algebra $\frk{co}$ --- the {\bf conformal algebra} of $V$. As a real vector space, we have
$$
\frk{co}:=V^*\oplus \frk{str}\oplus V.
$$ 
By writing $z\in V$ as $X_z$, $\langle w\mid\; \rangle\in V^*$ as $Y_w$, the commutation relations can be written as follow:
for $u$, $v$, $z$, $w$ in $V$, 
\begin{eqnarray}\label{TKKRel}
\begin{matrix}[X_u, X_v] =0, \quad [Y_u, Y_v]=0, \quad [X_u,
Y_v] = -2S_{uv},\cr\\ [S_{uv},
X_z]=X_{\{uvz\}}, \quad [S_{uv}, Y_z]=-Y_{\{vuz\}},\cr\\
[S_{uv}, S_{zw}] = S_{\{uvz\}w}-S_{z\{vuw\}}.
\end{matrix}
\end{eqnarray}
Note that, when the Jordan algebra is $\Gamma(3)$: $\mathbb R\oplus \mathbb R^3$ (a linear subspace of the real Clifford algebra $\mathrm{Cl}(\mathbb R^3, \mbox{dot product})$) with the product being the symmetrized Clifford multiplication, we have $\frk{co}=\frk{so}(2,4)$ --- the conformal algebra of the Minkowski space, and $\frk{str}=\frk{so}(1,3)\oplus \bb R$.

By definition, the universal enveloping algebra of $\frk{co}$ is called the {\bf TKK algebra}. The simply connected real Lie group with $\frk{co}$ as its Lie algebra is called the {\bf conformal group} and is denoted by $\mathrm{Co}$.

\section{The Universal Kepler Problem}
This section is the core of this article. The novel idea introduced here came from the author's realization that the Kepler problem can be reformulated in terms of $\Gamma(3)$ and the further realization that $\Gamma(3)$ can be replaced by any Euclidean Jordan algebra.

Hereafter we shall assume that $V$ is a simple Euclidean Jordan algebra. To introduce the universal Kepler problem associated with $V$, one needs first to complexify the TKK algebra and then formally invert the element $Y_e$, here $e$ is the unit element of $V$. With that done, one introduce
the {\bf universal hamiltonian}
\begin{eqnarray}\label{universalH}
H:={1\over 2} Y_e^{-1} X_e+\mathrm{i}Y_e^{-1}
\end{eqnarray} where $Y_e^{-1}$ is the formal inverse of $Y_e$, and $\mathrm i$ is the unit for imaginary numbers. 
 
Next, we introduce the {\bf universal Laplace-Runge-Lenz vector}
\begin{eqnarray}
A_u:=\mathrm{i}Y_e^{-1}[L_u, Y_e^2H]
\end{eqnarray} where $u$ is an element of $V$ and $[\;,\;]$ denotes the commutator. So 
\begin{eqnarray}A_u={\mathrm{i}\over 2}X_u-\mathrm{i}Y_uH  ={\mathrm{i}\over 2}(X_u-Y_uY_e^{-1}X_e)+Y_uY_e^{-1}.
\end{eqnarray}

Finally, we introduce the {\bf universal angular momentum}
\begin{eqnarray}
L_{u,v}:=[L_u, L_v]
\end{eqnarray} where $u, v$ are elements of $V$. We are now ready to state
\begin{theorem}\label{Main}
For any $u$, $v$, $z$ and $w$ in $V$, the following commutation relations
\begin{eqnarray}
\begin{array}{ll} 
[L_{u,v}, H] =  0\; ,& [A_u, H] = 0 \cr
[L_{u,v},  L_{z, w}] =  L_{[L_u, L_v]z,w}+ L_{z, [L_u, L_v]w} &\cr
[L_{u,v}, A_z] = A_{[L_u, L_v]z}\;,\ & [A_u, A_v] =
-2H L_{u,v}\;.
\end{array}
\end{eqnarray}
hold as identities in the resulting algebra obtained from complexifing the TKK algebra and formally inverting the element $Y_e$. 
\end{theorem}
\begin{proof}
Eq. (\ref{TKKRel}) implies that
$$
[L_{u,v},   X_z]=   X_{[L_u, L_v]z}, \quad [L_{u,v},  Y_z]=   Y_{[L_u, L_v]z},\quad [L_{u,v},  L_z] =  L_{[L_u, L_v]z}.
$$
In particular, we have $[L_{u,v},  X_e]=[L_{u,v},  Y_e]=0$. Therefore, $[L_{u, v}, H]=0$,
\begin{eqnarray}
[L_{u, v}, A_z] &=& \mathrm{i}Y_e^{-1}[L_{u,v}, [ L_z, Y_e^2H]]\cr
&=& \mathrm{i}Y_e^{-1}\left([[L_{u,v},  L_z], Y_e^2H]+[L_z, [L_{u,v}, Y_e^2H]]\right)\cr
&=&  \mathrm{i}Y_e^{-1}[L_{[L_u, L_v]z}, Y_e^2H]\cr
&=& A_{[L_u, L_v]z},\nonumber
\end{eqnarray}
and
\begin{eqnarray}
[L_{u, v}, L_{z, w}] &=&  [L_{u,v}, [ L_z, L_w]]\cr
&=& [[L_{u,v},  L_z],L_w]+ [ L_z, [L_{u,v},  L_w]]\cr
&=& [  L_{[L_u, L_v]z}, L_w]+ [ L_z, L_{[L_u, L_v]w}]\cr
&=&L_{[L_u, L_v]z, w}+ L_{z,[L_u, L_v]w}.\nonumber
\end{eqnarray}
Since $H=Y_e^{-1}({1\over 2}  X_e+\mathrm{i})$, we have
\begin{eqnarray}
[A_u, H] &=&  [A_u,Y_e^{-1}]({1\over 2}  X_e+\mathrm{i})+Y_e^{-1}[A_u, {1\over 2}  X_e+\mathrm{i}]\cr
&=&  -Y_e^{-1}[A_u,Y_e]Y_e^{-1} ({1\over 2}  X_e+\mathrm{i})+Y_e^{-1} [A_u, {1\over 2} X_e]\cr
&=&  -Y_e^{-1}[A_u,Y_e]H-Y_e^{-1}\left[{\mathrm{i}\over 2}X_u-\mathrm{i}Y_uH, {1\over 2}X_e\right]\cr
&=&  -Y_e^{-1}[A_u,Y_e]H-Y_e^{-1}\left[-\mathrm{i}Y_uY_e^{-1}, {1\over 2}X_e\right]Y_eH\cr
&=&  -Y_e^{-1}[A_u,Y_e]H+\mathrm{i}Y_e^{-1}\left( [Y_u, {1\over 2} X_e]-Y_uY_e^{-1}[Y_e, {1\over 2}X_e]\right) H\cr
&=&   -Y_e^{-1}[{\mathrm{i}\over 2}(X_u-Y_uY_e^{-1}X_e)+Y_uY_e^{-1},Y_e]H
+\mathrm{i}Y_e^{-1}\left( L_u-Y_uY_e^{-1} L_e\right) H\cr
&=&0.\nonumber
\end{eqnarray}
Since $A_u= {\mathrm{i}\over 2}X_u-\mathrm{i}Y_uH $, we have
\begin{eqnarray}
[A_u, A_v] &=& [{\mathrm{i}\over 2} X_u, -\mathrm{i}Y_v H ]-[{\mathrm{i}\over 2} X_v,  -\mathrm{i}Y_v H]-[Y_uH,Y_v H]\cr
&=& [{1\over 2} X_u, Y_vH ]-Y_u [H, Y_v] H-<u\leftrightarrow v>\cr
&=& [{1\over 2} X_u, Y_v] H +Y_v[{1\over 2} X_u,  H] -Y_u [H, Y_v] H -<u\leftrightarrow v>\cr
&=& - S_{uv} H -Y_vY_e^{-1}[{1\over 2} X_u,  Y_e] H-Y_uY_e^{-1} [{1\over 2} X_e, Y_v] H\cr && -<u\leftrightarrow v>\cr
&=& -S_{uv} H +Y_vY_e^{-1} L_u H+Y_uY_e^{-1} L_v H -<u\leftrightarrow v>\cr
&=& -2L_{u,v} H =-2HL_{u,v}.\nonumber
\end{eqnarray}
Here $<u\leftrightarrow v>$ means a term obtained from its immediate predecessor by switching $u$ with $v$.

\end{proof}
Because of Theorem \ref{Main} we say that the universal hamiltonian $H$ in Eq. \eqref{universalH} defines the {\bf universal Kepler problem} associated with the simple euclidean Jordan algebra $V$.   

\section{Concrete Realizations}
A concrete realization of the TKK algebra (equivalently conformal algebra) yields a concrete realization for the universal Kepler problem, i.e., a concrete model which resembles the Kepler problem. Of course, certain condition on the concrete realization of the TKK algebra must be satisfied. For example, $H$, $A_u$ and $L_{u, v}$ must be represented as real functions in a (classical) Poisson realization  and as self-adjoint operators in a (quantum) operator realization.

\subsection{Poisson realizations}
We are only interested in a Poisson realization of the TKK algebra on a Poisson manifold in which $S_{uv}$, $X_z$, $Y_w$ are all realized as real functions $\mathcal S_{uv}$, $\mathcal X_z$, $\mathcal Y_w$ respectively, with $\mathcal X_e$ and $\mathcal Y_e$ being both everywhere positive. Then, $H$, $A_u$ and $L_{u,v}$ can be realized as
\begin{eqnarray}\label{universalHcla}
\mathcal H={{1\over 2} \mathcal X_e-1\over \mathcal Y_e},\quad \mathcal A_u:={\{\mathcal L_u, \mathcal Y_e^2\mathcal H\}\over \mathcal Y_e}, \quad \mathcal L_{u,v}:=\{\mathcal L_u, \mathcal L_v\}
\end{eqnarray} respectively. Note that
\begin{eqnarray}\mathcal A_u={1\over 2}\left(\mathcal X_u-\mathcal Y_u{\mathcal X_e\over \mathcal Y_e}\right)+{\mathcal Y_u\over \mathcal Y_e}.
\end{eqnarray}
Theorem \ref{Main} obviously holds under the following substitutions:
$$
[, ]\to\{,\}, \quad H\to \mathcal H, \quad A_u\to \mathcal A_u, \quad L_{u,v}\to {\mathcal L}_{u, v}.$$

\subsubsection{Examples (without magnetic charge)} As is well-known, the total cotangent space $T^*V$ is a natural symplectic space.  By virtue of the invariant inner product on $V$, one can identify $T^*V$ with the total tangent space $TV$, then $TV$ becomes a symplectic space. The tangent bundle of $V$ has a natural trivialization, with respect to which, one can denote an element of $TV$ by $(x,\pi)$.  We fix an orthonormal basis $\{e_\alpha\}$ for $V$ so that we can write $x=x^\alpha e_\alpha$ and $\pi=\pi^\alpha e_\alpha$. Then the basic Poisson bracket relations on $TV$ are 
$$\{x^\alpha, \pi^\beta\}=\delta^{\alpha\beta}, \quad\{x^\alpha, x^\beta\}=0, \quad\{\pi^\alpha, \pi^\beta\}=0.$$
One can check that real functions
\begin{eqnarray}\label{MomentMap}
{\mathcal S}_{uv} :=\langle S_{uv}(x)\mid \pi\rangle, \quad {\mathcal X}_u: = \langle x\mid \{\pi u\pi \}\rangle, \quad {\mathcal Y}_v: =\langle x\mid v\rangle
\end{eqnarray} yield a Poisson realization on $TV$ of $S_{uv}$, $X_z$, $Y_w$ respectively. However, neither $\mathcal X_e$ nor $\mathcal Y_e$ is positive on $TV$. To salvage this Poisson realization, we restrict the Poisson realization to certain sub-symplectic manifolds of $TV$, for example, $T{\mathcal C}_r$ where $\mathcal C_r$ is the set of rank $r$ semi-positive elements of $V$, with $r$ being a positive integer less than or equal to the rank of $V$. It is an observation in Ref. \cite{meng11a} that $\mathcal X_e$ and $\mathcal Y_e$ are both positive on $T{\mathcal C}_r$ so that $\mathcal H$ defines a generalized Kepler problem; in fact, if $V=\Gamma(3)$ and $r=1$, then $\mathcal H$ defines the Kepler problem. 

\subsubsection{Examples (with magnetic charges)}
Let ${\mathbb R}^{2k+1}_*={\mathbb R}^{2k+1}\setminus\{\vec 0\}$ ($k\ge 1$) and $\pi$: ${\mathbb R}^{2k+1}_*\to \mathrm{S}^{2k}$ be the map sending $\vec r\in {\mathbb R}^{2k+1}_*$ to ${\vec r\over |\vec r|}\in \mathrm{S}^{2k}$.  Denote by $P\to {\mathbb R}^{2k+1}_*$ the pullback by $\pi$ of the canonical principal $\mathrm{SO}(2k)$-bundle $\mathrm{SO}(2k+1)\to \mathrm{S}^{2k} $. Let $E\to {\mathbb R}^{2k+1}_*$ be the associated co-adjoint bundle for $P\to {\mathbb R}^{2k+1}_*$ and $E^\sharp\to T^*{\mathbb R}^{2k+1}_*$ be the pullback bundle under the cotangent bundle projection map $T^*{\mathbb R}^{2k+1}_*\to {\mathbb R}^{2k+1}_*$. It is a fact that the canonical connection on $\mathrm{SO}(2k+1)\to \mathrm{S}^{2k} $ turns $E^\sharp$ into a Poisson manifold. It has been shown in Ref. \cite{meng12} that the real Lie algebra $\mathfrak{so}(2, 2k+2)$ --- the conformal Lie algebra of the Jordan algebra $\Gamma(2k+1):=\mathbb R\oplus \mathbb R^{2k+1}$ ---  has a Poisson realization on certain symplectic leaves of $E^\sharp$, and each of these Poisson realizations yields a magnetized Kepler problem in dimension $2k+1$. For more details, please consult Ref. \cite{Meng'}.

\subsection{Operator realizations}
Throughout this section we assume that the Jordan algebra is $\bb R$. Then the symmetric cone is ${\bb R}_+:=(0,\infty)$ and $\mr{Co}=\widetilde{\mr{SL}}(2, \bb R)$ --- the universal cover of $\mr{SL}(2, \bb R)$. A point in $\bb R$ is denoted by $x$, and the Lebesgue measure on $\bb R$ is denoted by $\mathrm{d}x$.
The conformal algebra is $\frk{sl}(2, \bb R)$, with generators $S:=S_{ee}$, $X:=X_e$ and $Y:=Y_e$ and commutation relations
$$
[S, X]=X, \quad [S, Y]=-Y, \quad [X, Y]=-2S.
$$
These generators can be realized as linear operators on $L^2\left(\bb R_+, {1\over x}\,  \mathrm{d}x\right)$ as follows:
$$
S\to\tilde S:=-x{ \mathrm{d}\over  \mathrm{d}x}, \quad X\to\tilde X(\nu):=\mathrm{i}\left(x{ \mathrm{d}^2\over  \mathrm{d}x^2}+{{\nu\over 2}(1-{\nu\over 2})\over x}\right), \quad Y\to \tilde Y:=-\mathrm{i}x.
$$   
Here, $\nu$ is a complex parameter whose range is to be determined. Note that, we must specify a common dense domain of definition for these operators. This common domain $\tilde D_\nu$ is defined to be
\begin{eqnarray}
\left\{x^{\nu\over 2}e^{-x}p(x)\mid p(x) \in \bb C[x]\right\}.\nonumber
\end{eqnarray}
For $\nu\in (0, \infty)$ and only for such an $\nu$, $x^{\nu-1}e^{-2x}\, \mathrm{d}x$ is a finite positive measure on $\bb R_+$. Therefore, for such and only for such a $\nu$, $\bb C[x]$ is dense in $L^2\left(\bb R_+, x^{\nu-1}e^{-2x}\, \mathrm{d}x\right)$, or equivalently, $\tilde D_\nu$ is dense in $L^2\left(\bb R_+, {1\over x}\,  \mathrm{d}x\right)$.

It is not hard to see that operators $\tilde S$, $\tilde X(\nu)$ and $\tilde Y$ are all anti-hermitian operators on $\tilde D_\nu$ when $\nu\in(0, \infty)$. Therefore, \begin{center}\emph{for each $\nu\in(0, \infty)$, $\tilde D_\nu$ is a unitary module $\pi_\nu$ for $\frk{sl}(2, \bb R)$;}\end{center} moreover\footnote{It appears that $\pi_\nu=\pi_{2-\nu}$ for $\nu\in (0, 2)$, but that is not true, because
$\tilde D_\nu\neq \tilde D_{2-\nu}$.}, $\pi_{\nu_1}\not\cong \pi_{\nu_2}$ if $\nu_1\neq \nu_2$.

Let $E_\pm={\mathrm{i}\over 2}(\tilde X(\nu)-\tilde Y)\mp \tilde S$, $h={\mathrm{i}\over 2}(\tilde X(\nu)+\tilde Y)$. Suppose that $h(\psi_s)=s\psi_s$ and $E_-(\psi_s)=0$, then $\psi_s\propto x^s e^{-x}$ with $s={\nu/2}$.  Therefore,
$x^{\nu\over 2} e^{-x}$ is a lowest weight state for $\pi_\nu$. After a little play with algebra, one can show that
$\pi_\nu$ is the lowest weight module for $\frk{sl}(2, \bb R)$, in fact, a unitary lowest weight $(\frk{g}, \mr{K})$-modules, where $\frk{g}=\frk{sl}(2, \bb R)$, $\mr{K}\cong \bb R$ is the simply connected abelian group whose Lie algebra is generated by $X+Y$. Since $\tilde D_\nu$ is dense in $L^2\left(\bb R_+, {1\over x}\,  \mathrm{d}x\right)$, via integration, we obtain a unitary lowest weight  representation (also denoted by $\pi_\nu$) of $\widetilde{\mr{SL}}(2, \bb R)$ on $L^2\left(\bb R_+, {1\over x} \, \mathrm{d}x\right)$. In view of the classification theorem for unitary lowest weight modules in Ref. \cite{EHW82}, these $\pi_\nu$ exhaust all nontrivial unitary lowest weight representations of $\widetilde{\mr{SL}}(2, \bb R)$.

\vskip 10pt
Combining with the result in the previous section, for each $\nu\in (0, \infty)$, there is a generalized (quantum) Kepler problem whose hamiltonian is
$$
\tilde H(\nu)=-{1\over 2}{ \mathrm{d}^2\over  \mathrm{d}x^2}+{{\nu\over 2}({\nu\over 2}-1)\over 2x^2}-{1\over x}.
$$
However, the Laplace-Runge-Lenz vector is trivial: $A_u=u$.  It appears that $\tilde H(\nu)=\tilde H(2-\nu)$ for $\nu\in (0, 2)$, but that is not true, because
$\tilde H(\nu)$ and $\tilde H(2-\nu)$ have {\it different} domains of definition\footnote{This can be verified from the requirements that $\tilde H(\nu)$ be hermitian with respect to inner product
$$
(f,g)\mapsto \int_{\bb R_+}\bar f g\, \mathrm{d}x,
$$ and its domain of  definition contain $x^{\nu\over 2}e^{-{2x\over \nu}}$.}.

The bound state spectrum for $\tilde H(\nu)$ is $$\left\{-{1/2\over (I+\nu/2)^2}\mid I=0, 1, \ldots\right\};$$ moreover, being a closed subspace of $L^2(\bb R_+, \mathrm{d}x)$, the Hilbert space of bound states for $\tilde H(\nu)$ is isometric to $L^2\left(\bb R_+, {1\over x}\, \mathrm{d}x\right)$ via an analogue of the twisting map $\tau$ introduced in the proof of Theorem 5 in Ref. \cite{meng09} and hence provides another realization for $\pi_\nu$. 

\section*{Acknowledgements}
The author was supported by Qiu Shi Science and Technologies Foundation while he was a member at the Institute for Advanced Studies in the cadmic year 2010-2011. He was also supported by the Hong Hong Research Grants Council under RGC Project No. 16304014.

\end{document}